\documentclass[aps,prl,preprint]{revtex4}

\usepackage{ifpdf}
\ifpdf
\fi

\usepackage[utf8]{inputenc}

\usepackage{graphicx}
\usepackage{epstopdf}
\usepackage{dcolumn}
\usepackage{bm}

\usepackage{SIunits}
\usepackage{subfigure}
\usepackage{hyperref}
\usepackage{amsmath} % eqref and other goodies

\include{preamble}

\begin{document}
%opening

\title{Impact of Detector Solenoid on the CLIC Luminosity Performance}
\author{Y.~Inntjore~Levinsen}
\email{yngve.inntjore.levinsen@cern.ch}
\affiliation{CERN}
\author{B.~Dalena\footnote{Now at CEA/SACLAY, DSM/Irfu/SACM F-91191 Gif-sur-Yvette France.}}
\affiliation{CERN}
\author{R.~Tom\'{a}s}
\affiliation{CERN}
\author{D.~Schulte}
\affiliation{CERN}

\begin{abstract}

In order to obtain the necessary luminosity with a reasonable amount of beam power, the Compact LInear
Collider (CLIC) design includes an unprecedented collision beam size of
$\sigma_y$ = \unit{1}{\nano\meter} vertically and $\sigma_x$ = \unit{45}{\nano\meter} horizontally.
With exceptionally small and flat beams, the luminosity can be significantly degraded due to the combination of the
experimental solenoid field and a large crossing angle.
The two main effects reducing the luminosity are y-x$'$-coupling and increase of vertical dispersion.
Additionally, Incoherent Synchrotron Radiation (ISR) from the orbit deflection created by the solenoid field increases the beam emittance 
and result in unrecoverable luminosity degradation.

A novel approach to evaluate the ISR effect from a realistic solenoid field without knowledge of the full compensation of the geometric 
aberrations is presented. This approach is confirmed by a detailed study of the correction techniques to compensate the beam optics 
distortions. The unrecoverable luminosity loss due to ISR for CLIC at 3~TeV has been evaluated, and found to be around 4~\% to 5~\% for the 
solenoid design under study.

\end{abstract}

\maketitle

\section{Introduction}

CLIC is an accelerator design based on normal conducting components. In order to obtain the required luminosity with reasonable power 
consumption, short bunch separation (\unit{0.5}{\nano\second}) and small $\beta^*$ are needed.
A post-collision beamline for the spent beam and the main beam-beam products is necessary, which requires a large crossing angle of around
\unit{20}{\milli\rad} \cite{Schulte01:crossangleCLIC}. A large crossing angle is also required to mitigate the effects of the 
parasitic bunch collisions between the incoming and outgoing beam, but is not the limiting factor for CLIC \cite{Schulte01:crossangleCLIC}.

Relevant parameters for the CLIC Beam Delivery System (BDS) are shown in Table~\ref{tab:clic_bds_param}. A detailed overview of the CLIC 
BDS can be found in  \cite[Ch. 3.5]{CLIC_CDR_V1}. The CLIC final focus system has strict tolerances, and the BDS is optimised taking higher 
order terms into consideration~\cite{RaimSery01:NovelFinalFocusDesigFuturLineaColli, TomasGarcia06:Nonloptibeamline, 
Marin12:PhDDesighigheorderoptimFinalFocusSysteLineaColli, Marin14:DesighighorderoptimAccelTestFacillatti, 
BarrTomaMari13:LuminstuditravewaistregimCompaLineaColli}. Compensation of beam distortions in the BDS, such as static and dynamic 
misalignments, has proven to be quite challenging~\cite{Dalena12:BDStuninluminmonitCLIC, CLIC_CDR_V1}.

\begin{table}
\centering
\caption{CLIC BDS parameters \cite{CLIC_CDR_V1}. Peak luminosity is defined as the luminosity in the 1~\% energy peak.}
  \begin{tabular}{lc}
  \hline
  Parameter          & Value \\
  \hline
  Maximum beam energy   & \unit{1.5}{\tera e\volt} \\
  L$^*$              & \unit{3.5}{\meter} \\
  $\beta^*$ (x/y)    & \unit{10/0.07}{\milli\meter} \\
  Crossing angle     & \unit{20}{\micro\rad} \\
  IP beam size (x/y) & \unit{45/1}{\nano\meter} \\
  IP beam divergence (x/y) & \unit{7.7/10.3}{\micro\rad} \\
  Bunch length       & \unit{44}{\micro\meter} \\
  Nominal peak luminosity       & \unit{2.5\times 10^{34}}{\centi\meter^{-2}\second^{-1}} \\
  \hline
  \end{tabular}
\label{tab:clic_bds_param}
\end{table}

\begin{figure}
 \centering
 \includegraphics[width=0.5\textwidth]{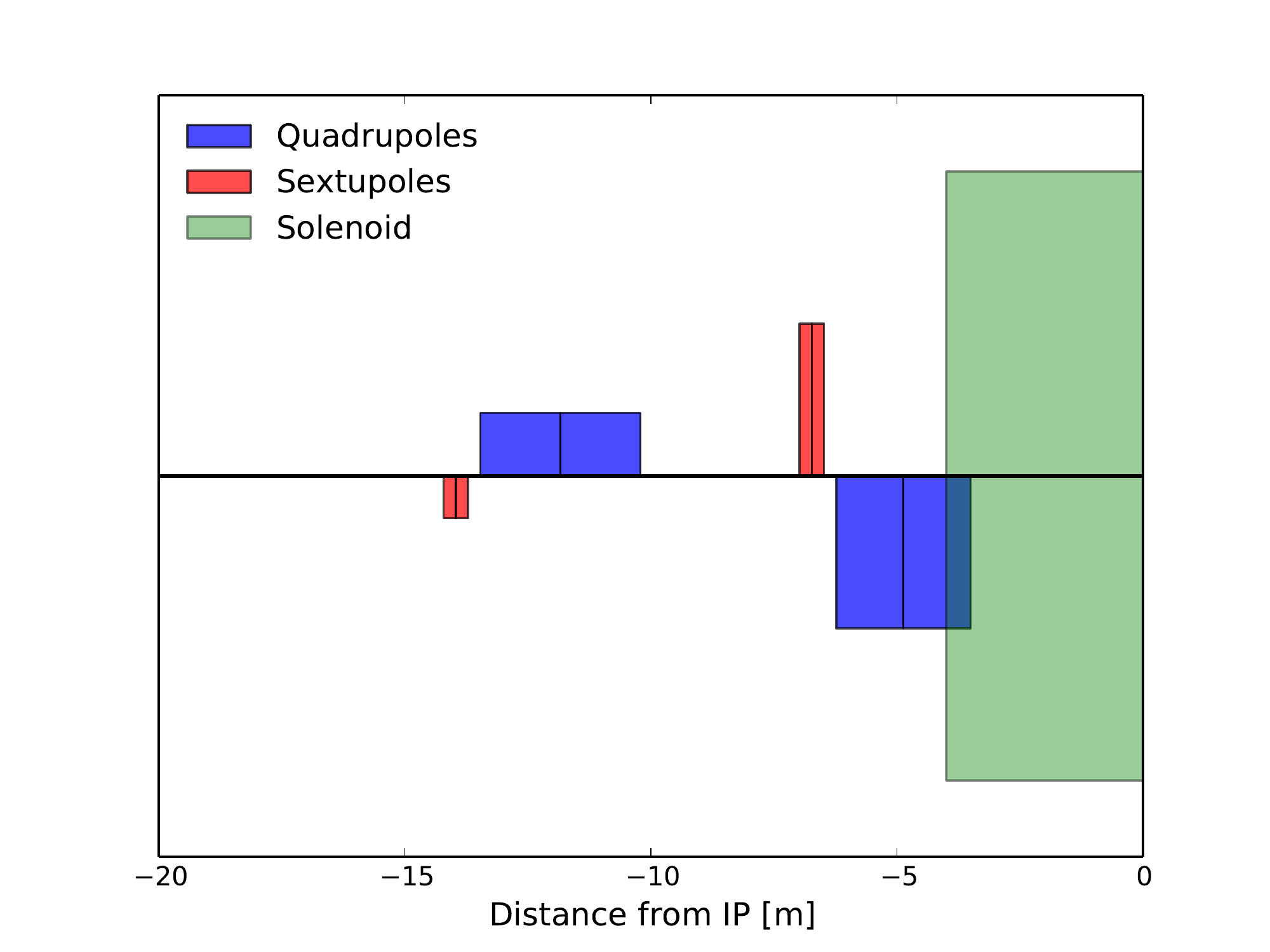}
 % clic_ff_lattice.pdf: 576x432 pixel, 72dpi, 20.32x15.24 cm, bb=0 0 576 432
 \caption{The final 20~m of the CLIC final focus system. The 4~m long experimental solenoid is marked in green. The final doublet 
quadrupoles are marked in blue, and the sextupoles in red. The height of the bars indicate their relative 
strength and polarity.}
 \label{fig:clic_ff_lattice}
\end{figure}

In Fig.~\ref{fig:clic_ff_lattice}, the final 20~m of the CLIC BDS lattice is shown. The residual field from an experimental solenoid 
typically extends 10-15~m away from the interaction point (IP), depending on shielding and solenoid design.  L$^*$ is the 
distance from the IP to the closest 
focusing magnet, which is the QD0 for CLIC. Due to the short L$^*$ required to reach the luminosity target, the main solenoid field 
overlaps with the 
last final focus magnets, which enhances the optical distortions at the IP~\cite{Nosochkov05:Compdetesoleeffebeamsizelinecoll}. The QD0 is 
partly inside the experimental solenoid. Special care has to be taken to make sure the interplay between the 
solenoid field and the magnet field is minimised.

A solenoid will in general have a radial field component on any charged particle off the solenoid centreline, with a maximum
around the entrance of the solenoid. This is the region of maximum $\beta$-function in a linear collider, and the
beams are more sensitive to small errors.
With a horizontal crossing angle, the horizontal solenoid field component will be larger than the
vertical one, resulting in a strong vertical orbit displacement.
In CLIC this orbit offset is typically on the order of \unit{10}{\micro\meter}, for a solenoid field of \unit{4-5}{\tesla} and 
\unit{1.5}{\tera e\volt} beam energy. 
The displacement results in a large vertical dispersion at the interaction point (IP).
Furthermore, the beams in CLIC are exceptionally flat, which means that any coupling to the vertical plane significantly deteriorates the 
luminosity.

Particles with large angles at the IP have a large displacement from the beam orbit in the region close to the
last focusing magnet, where the radial solenoid field is strongest. Hence, the experimental solenoid introduces strong y-x$'$
coupling at the IP which must be corrected.

Due to the high beam energy in CLIC, there is a significant emission of synchrotron radiation as a result of the beam deflection in 
the solenoid region.
Earlier similar studies have shown an unrecoverable luminosity loss due to ISR of up to \unit{25}{\%}, depending on the detector 
solenoid design \cite{Dale10:ImpacExperSolenCLICLumin, CLIC_CDR_V1}.

The unrecoverable loss is an important concern for CLIC. Optical aberrations can be corrected in several ways; using the final focus 
magnets, adding skew quadrupoles, using an anti-solenoid~\cite{Nosochkov05:Compdetesoleeffebeamsizelinecoll}, dipole orbit 
corrector integrated into the experiment~\cite{Seryi06:IROptiDIDanti}, and longer L$^*$ \cite{Seryi08:NearIRFFdesigincluFDlongeLissue}. 

We present a new simulation approach which evaluates the effect of the ISR alone without the knowledge of the full compensation. This 
approach is verified with a semi-analytical approach, as well as a more time-consuming study where the full compensation is found.

For the latter study, the tuning methods described in~\cite{Dalena12:BDStuninluminmonitCLIC} are used to compensate for optical 
distortions introduced by the experimental solenoid field. A realistic design of the solenoid and anti-solenoid is used 
~\cite{Swobod09:CLICantisolenoid}. Similar correction schemes have been explored for e.g. the NLC 
\cite{Nosochkov05:Compdetesoleeffebeamsizelinecoll}, but at lower beam energies which means synchrotron radiation effects are less 
significant.

There are two problems with the full compensation study which are addressed with the new simulation approach. First of all, it is a 
computationally demanding procedure, requiring on the order of weeks of CPU time to get to the final result. Second, once the result is 
obtained, one does not know if the remaining luminosity loss is purely due to ISR, or if there are residual optical aberrations.

\section{Experimental Solenoid Field}

Two detectors will be running in a push-pull configuration in CLIC. In the conceptual design report it is foreseen that one detector will
follow the SiD design~\cite{Aiha09:SiDLetteInten}, while the second detector will have an ILD design~\cite{ILD10:ILDLetterIntent}. 
An important difference in the two magnet designs is the peak longitudinal field, which is \unit{4}{\tesla} for the ILD detector 
magnet, and \unit{5}{\tesla} in the SiD case. Nevertheless, previous studies have found that the luminosity loss from the two detector 
designs
is fairly similar due to the relative increase in stray fields from the ILD solenoid compared to the SiD solenoid
\cite{Dale10:ImpacExperSolenCLICLumin}.

\begin{figure}
 \centering
  \subfigure[~B$_\text{z}$]{
    \includegraphics[width=0.45\textwidth]{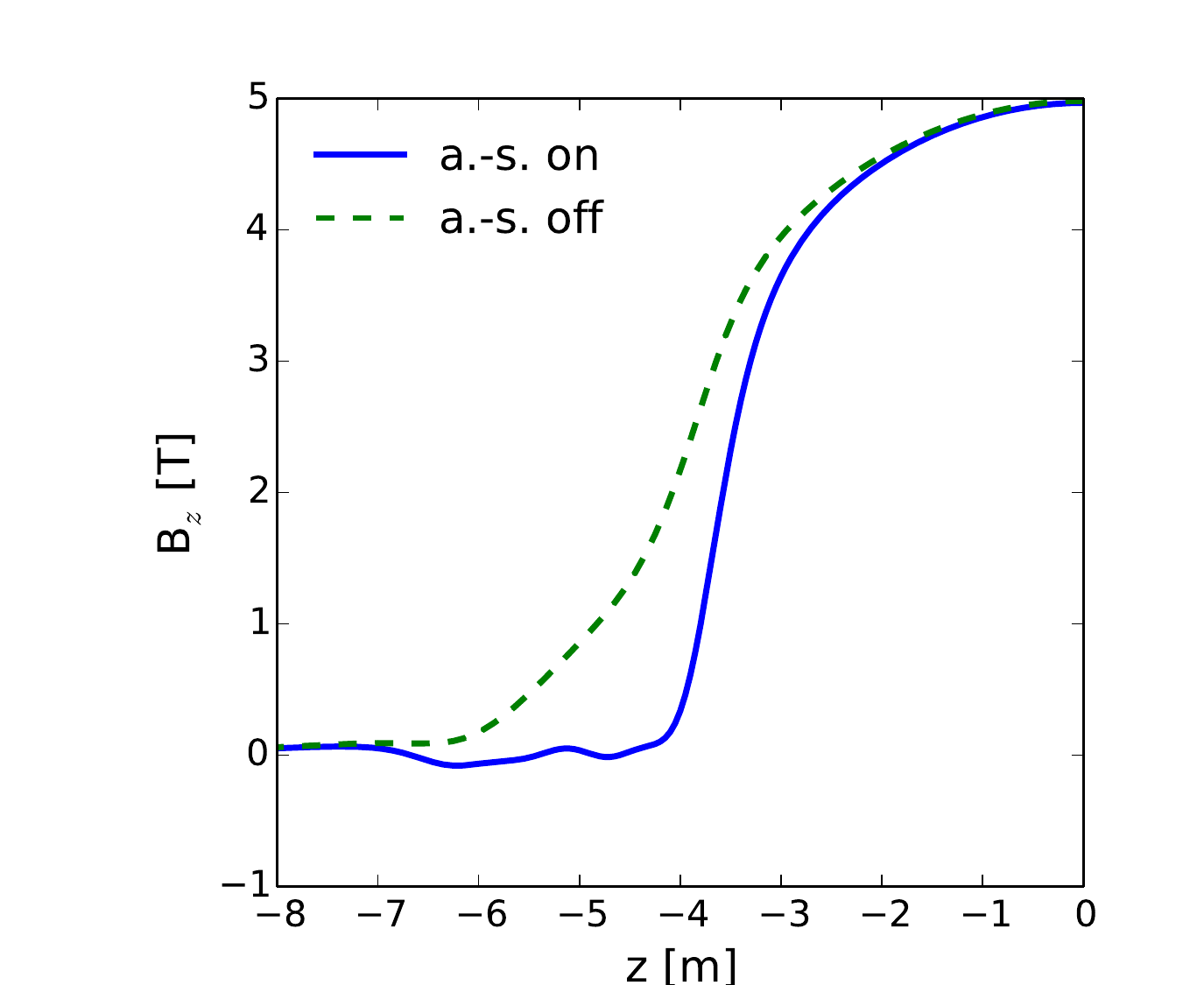}
  % field_map_bz.pdf: 288x216 pixel, 72dpi, 10.16x7.62 cm, bb=0 0 288 216
  }
  \subfigure[~B$_\text{r}$]{
    \includegraphics[width=0.45\textwidth]{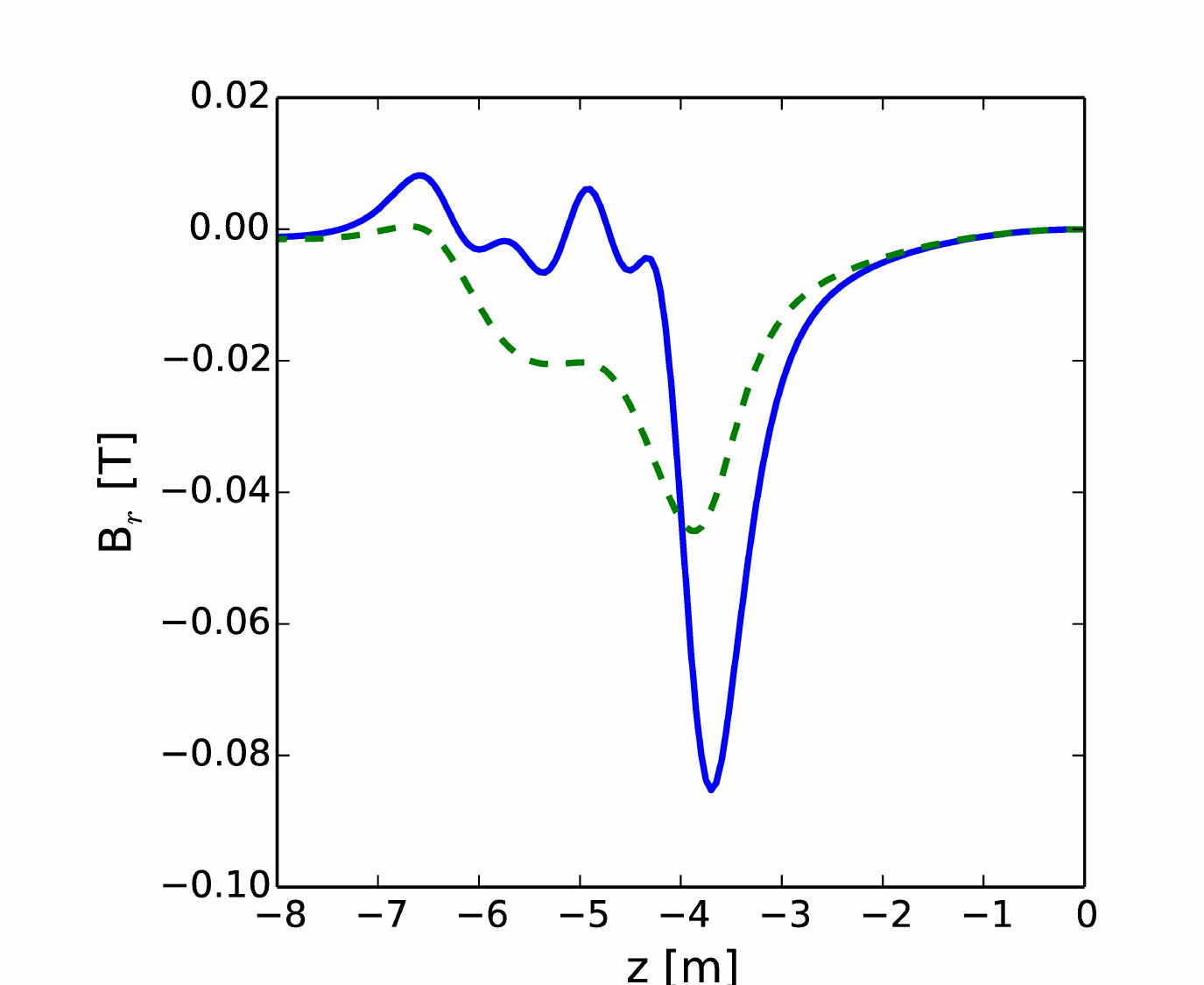}
  % field_map_br.pdf: 288x216 pixel, 72dpi, 10.16x7.62 cm, bb=0 0 288 216
  }
 \caption{The longitudinal (a) and radial (b) SiD solenoid field with (blue, solid) and without (green, dashed) anti-solenoid, along a
beamline with a \unit{10}{\micro\rad} inclination with respect to the solenoid axis. The QD0 entrance is at \unit{3.5}{\meter}, and the IP 
is at \unit{0}{\meter}.}
 \label{fig:sid_fieldmap}
\end{figure}

The longitudinal and radial fields along the beamline for the SiD detector magnet are shown in Fig.~\ref{fig:sid_fieldmap}, both with 
(blue, solid) and without (green, dashed) the anti-solenoid~\cite{Aiha09:SiDLetteInten}. For all detector designs currently considered, 
the anti-solenoid is foreseen to be integrated into the CLIC detector. This is in contrast to e.g. the International Linear 
Collider 
(ILC) \cite{13:ILCTechnDesigRepor}, 
where the anti-solenoid is integrated into the QD0 design \cite{Parker07:supermagneILCbeamdelivsyste}.
The anti-solenoid significantly reduces the longitudinal field 
inside the QD0, increasing the radial field at the entrance of the QD0 (at 3.5~m). This reduces the 
optical aberrations originating from the combination of the quadrupolar field of the QD0 and the solenoid stray fields 
\cite{Nosochkov05:Compdetesoleeffebeamsizelinecoll}.

In this paper we discuss only the L$^*$ = \unit{3.5}{\meter} lattice design. An increase of L$^*$ to around \unit{6-8}{\meter} has been
considered in order to have the QD0 outside the detector \cite{Seryi08:NearIRFFdesigincluFDlongeLissue,
Zamudi10:OptimCLIC500GeVFinalFocussystedesigifnew3TeVFinalFocussysteL=60M}. This would have the benefit of an improved mechanical stability 
of the QD0 since it would be attached to the tunnel ground/beamline. It would further reduce the interference between the QD0 and the 
experiment, both in terms of taking up physical space for detector components, and in terms of magnetic field interplay. The downside of a
longer L$^*$ is primarily reduced luminosity. A different L$^*$ is expected to significantly impact the result of the studies presented 
here.

\section{Semi-Analytical Approach}

The problem of evaluating the luminosity loss due to the detector solenoid can be divided into two parts.
The first part consists of the evaluation and the correction of the optical distortions, which should be possible to correct for by 
using the anti-solenoid and tuning knobs.
The second part, the ISR from the vertical orbit deflection, increases the beam emittance. This emittance increase cannot be compensated 
for, and can be considered a minimum luminosity loss for a given solenoid design.

From Refs.~\cite{ParkSery05:Compeeffecdetecsolenvertibeamorbitlineacolli, TeneIrwiRaub03:Beamdynaminterregiosolenlineacolliduetocrossangle},
we have the following estimate for the increase of vertical beam size due to synchrotron radiation
\begin{equation}
 \left(\Delta \sigma_y^{SR}\right)^2 = C_E\gamma^5 \int_0^\infty \frac{R_{36}^2(z)}{\left|\rho(z)\right|^3} \mathrm{d}z,
\end{equation}
where
\begin{equation}
 C_E = \frac{55}{24\sqrt(3)} r_e \lambda_e = 1.26 \times 10^{-27}.
\end{equation}
Here, $R_{36}$ is the transport matrix element $36$ for the given slice $\mathrm{d}z$ to the IP, $\gamma$ is the relativistic gamma, and 
$\rho(z)$ is the radius of curvature at $z$. $r_e$ and $\lambda_e$ are the electron classical and Compton radius respectively. The beam 
size increase should be added in quadrature to the initial beam size.

\begin{figure}
\centering
\includegraphics[width=0.5\textwidth]{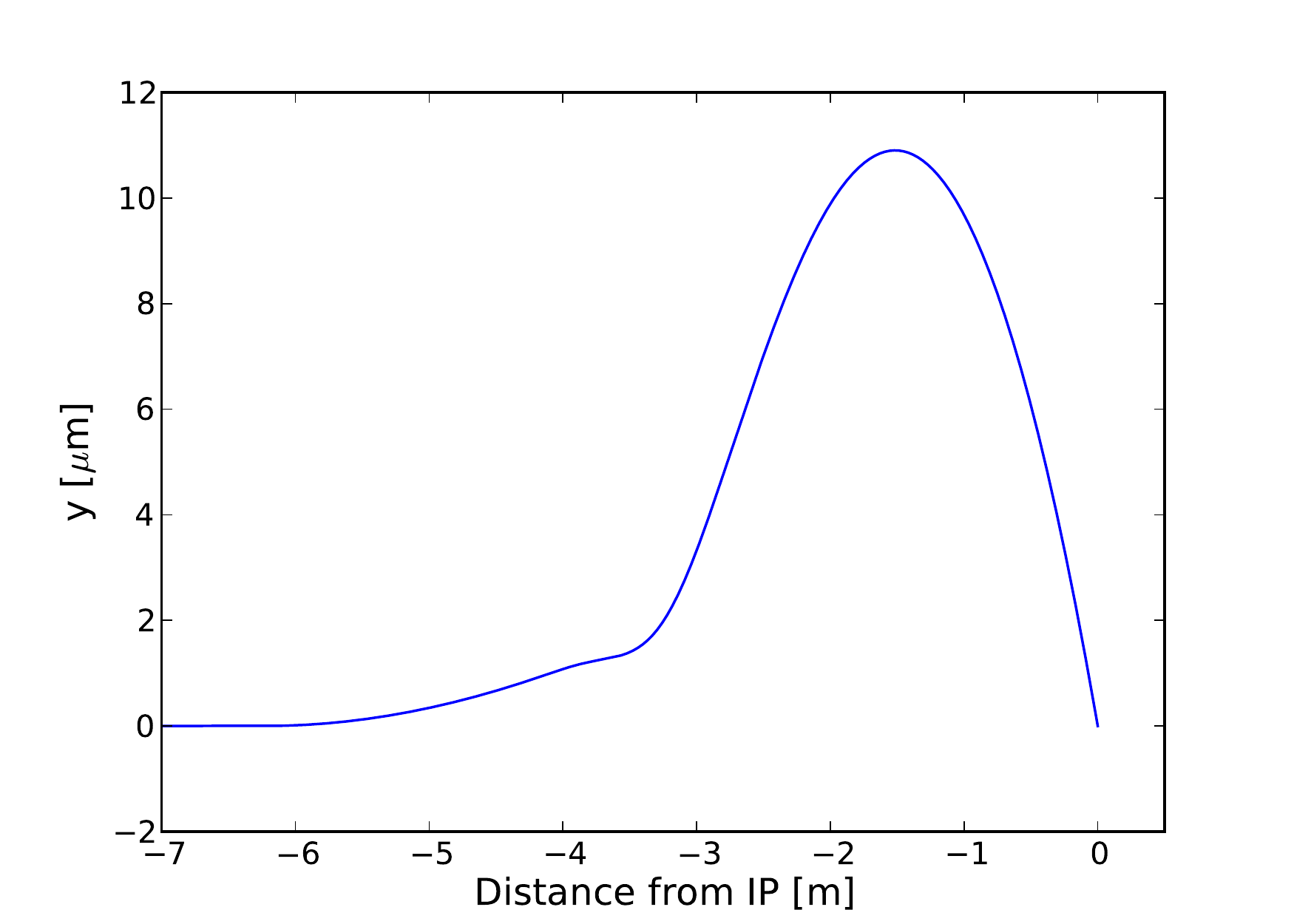}
% orbit_y.pdf: 504x360 pixel, 72dpi, 17.78x12.70 cm, bb=0 0 504 360
\caption{The vertical orbit of the last \unit{7}{\meter} before the IP given in \micro\meter. The QD0 is displaced vertically in order to 
get the orbit to end at y~=~0.}
\label{fig:vertical_orbit}
\end{figure}

We calculate the $R_{36}(z)$ by tracking backwards an off-momentum particle (\unit{+4}{\giga e\volt}) from the IP under the assumption that 
the dispersion at the IP 
is 0. $\rho(z)$ is calculated numerically from the orbit shown in Fig. \ref{fig:vertical_orbit}. The estimate from this analytical formula 
then gives us
\begin{equation}
 \Delta \sigma_y^{SR} = 0.36~\nano\meter,
\end{equation}
for the solenoid field map which has the anti-solenoid included.

% sqrt(1+0.36**2)/1-1
The initial core 1 sigma beam size is about \unit{1}{\nano\meter} in CLIC, which means an increase of 6.4~\%. If 
we assume that luminosity is inversely proportional to beam size, we get a luminosity loss of 6~\%. If we instead use the RMS vertical
beam size which is around \unit{1.3}{\nano\meter}, we get a relative increase to the beam size of around 3.7~\%. However, for the peak 
luminosity \footnote{The peak luminosity is defined as luminosity produced by collisions within 
1~\% of the energy peak. See discussion in e.g. \cite{CDRV2:PhyDetCLICLIConDesRep},~Ch.~2.1.1 } in particular, the core beam size is usually 
considered to be the more relevant parameter. This is an encouraging 
result, considering that up to 25~\% luminosity loss due to ISR caused by the solenoid was expected for the nominal CLIC machine 
\cite{CLIC_CDR_V1}.

Estimating luminosity loss only via beam size growth has a considerable level of uncertainty. The tails of the beam typically 
increase the RMS beam size, while not affecting luminosity as significantly. For this reason, we always use GUINEA-PIG 
\cite{Schult98:GUINEA-PIG} to simulate the luminosity in our tracking studies.

\section{Deterministic Approach}

\begin{figure}
 \centering
 \includegraphics[width=0.5\textwidth]{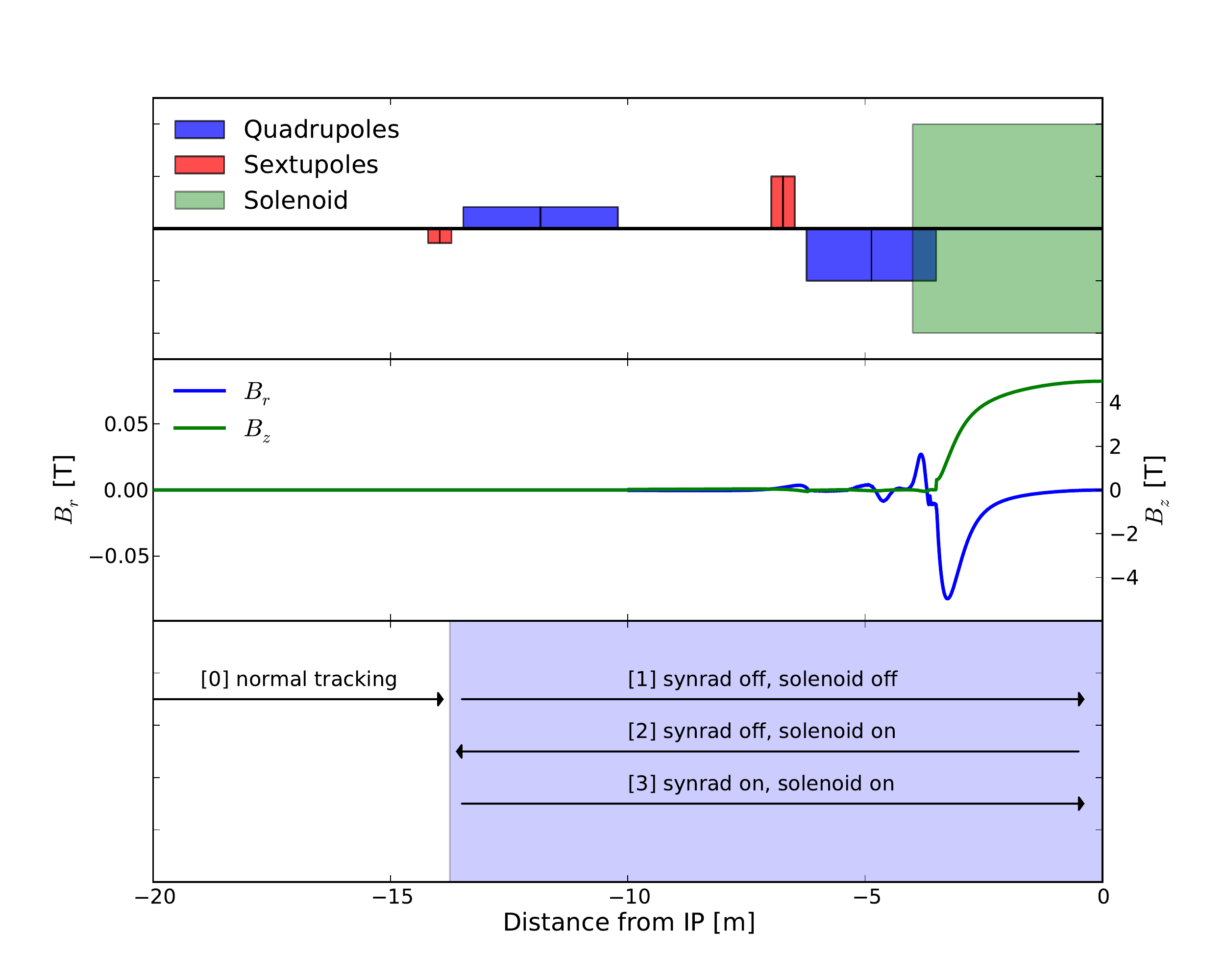}
 % deterministic_tracking.pdf: 720x576 pixel, 72dpi, 25.40x20.32 cm, bb=0 0 720 576
 \caption{Schematic overview of the last 20 metres of the final focus in CLIC in the upper third. The experimental solenoid (green) is 
overlapping the QD0 (blue). Sextupoles SD0 and SF1 in red. strengths and signs of quadrupoles/sextupoles are indicated by the size and 
direction of the bars. In the middle the simulated SiD solenoid field is shown. The radial field in blue with values on left side, and 
longitudinal field in green with values on right hand side. In the bottom plot the tracking procedure is visualised.}
 \label{fig:tracking_procedure}
\end{figure}

A novel simulation approach is proposed, which separately evaluates the losses from ISR alone, before the full 
compensation is known. The procedure to evaluate the luminosity loss due to ISR is described in the lower part of 
Fig.~\ref{fig:tracking_procedure}, where the tracking including the solenoid field is done using the new 4th order symplectic integrator 
described in the appendix.
The beam is first tracked forward without synchrotron radiation, and without the solenoid field present. 
This provides the optimal beam distribution at the interaction point. The
ideal IP beam distribution is tracked backwards through the beamline, 
with the solenoid field turned on but still without synchrotron radiation. 
The result is a beam distribution with a perfect compensation for the coupling introduced by the solenoid field. 
Finally, the synchrotron radiation is turned on, and the beam is tracked forward through the beamline.
The estimated luminosity is compared to a normal tracking of the beam without the solenoid field, but including ISR.

Using this approach we evaluated the simulated SiD field maps presented in Fig.~\ref{fig:sid_fieldmap}. 
The loss of peak luminosity due to ISR in the detector solenoid including the 
anti-solenoid is found to be ($4.1 \pm 0.2$)~\%, where the error bar is from the calculation of the luminosity in GUINEA-PIG.
This result compares well to the result from the semi-analytical calculation.

Without the anti-solenoid, we find a luminosity loss of around 5~\%.
1~\% of additional luminosity would not alone be enough to justify the installation of an anti-solenoid which significantly 
complicates the detector design. However, the most important purpose of the anti-solenoid is to protect the permanent magnet material in 
the QD0 and allow it to safely reach its high gradient.
The anti-solenoid also makes the compensation easier, as it decouples the solenoid field from the field inside the QD0 
\cite{Nosochkov05:Compdetesoleeffebeamsizelinecoll}.
With an alternative larger L$^*$ of 6~m or more \cite{Zamudi10:OptimCLIC500GeVFinalFocussystedesigifnew3TeVFinalFocussysteL=60M},
where the QD0 is outside the detector, the need for an anti-solenoid could be reassessed.

\section{Full Compensation}

\begin{figure*}
\centering
 \subfigure[~No solenoid]{
   \includegraphics[width=0.45\textwidth]{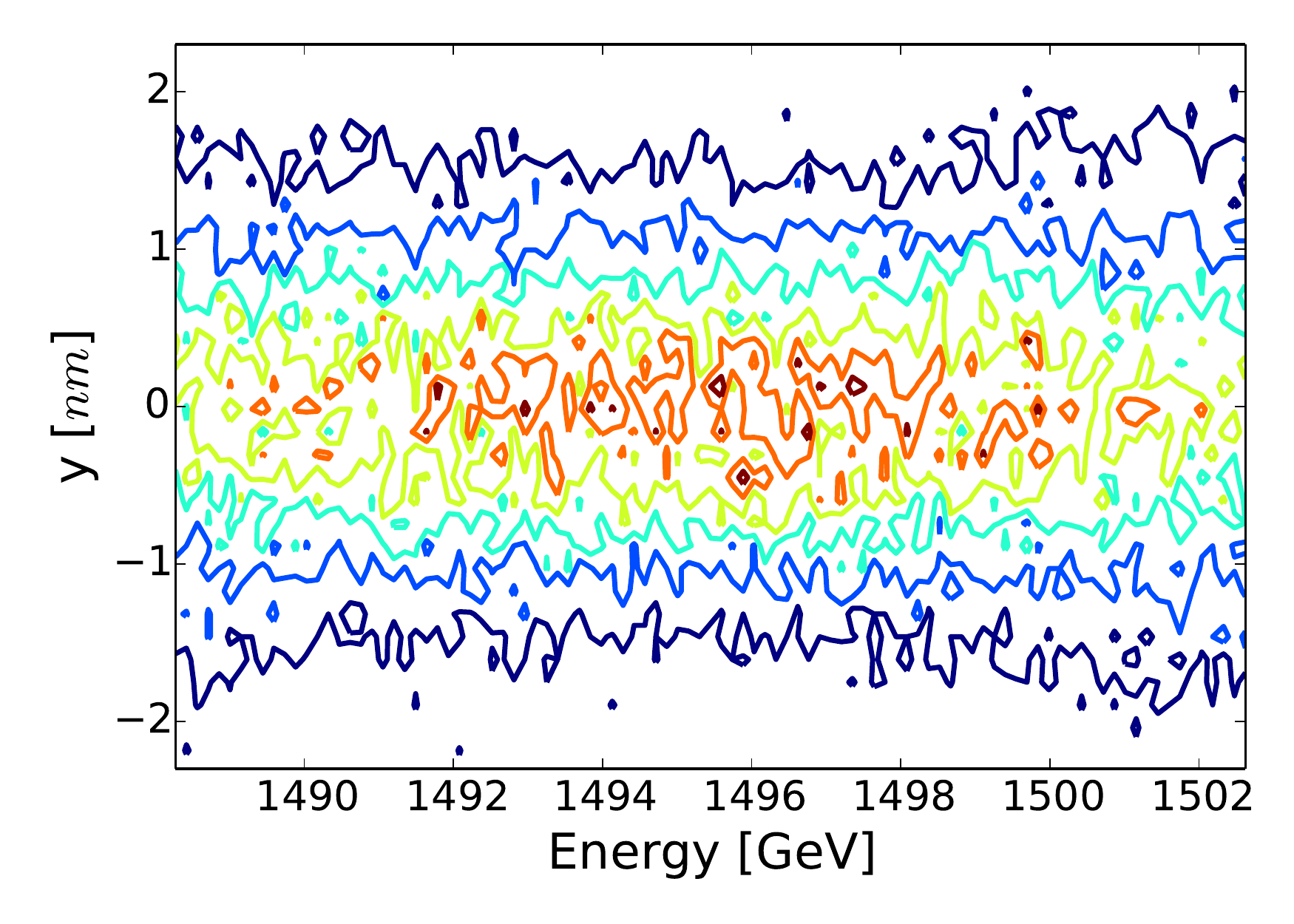}
   % contour_Ey.eps: 0x0 pixel, 300dpi, 0.00x0.00 cm, bb=54 216 558 576
   \label{fig:contour_Ey_nsol}
 }
 \subfigure[~With solenoid]{
   \includegraphics[width=0.45\textwidth]{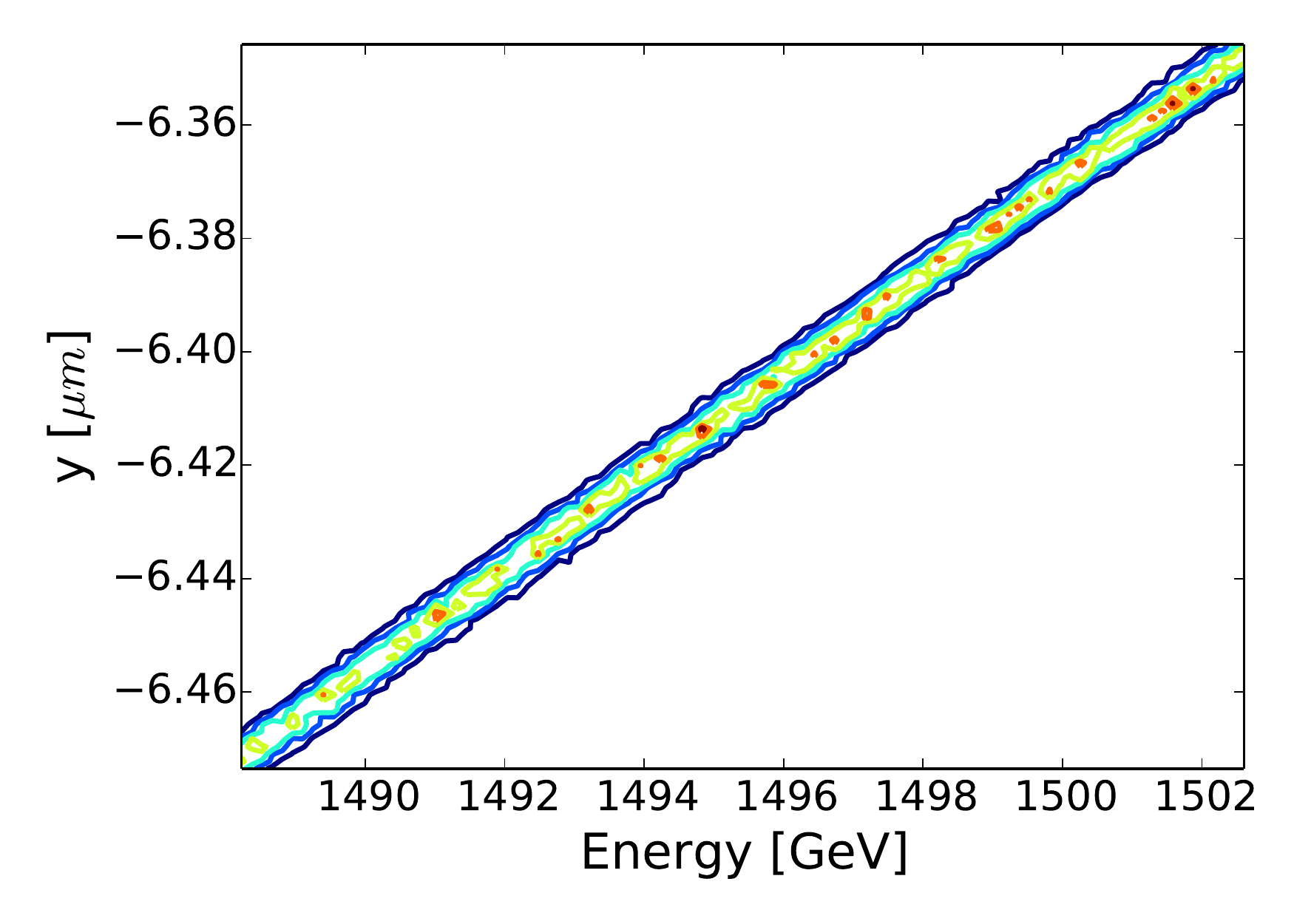}
   % contour_Ey.eps: 0x0 pixel, 300dpi, 0.00x0.00 cm, bb=54 216 558 576
   \label{fig:contour_Ey_wsol}
 }
 \subfigure[~With solenoid+anti-solenoid]{
   \includegraphics[width=0.45\textwidth]{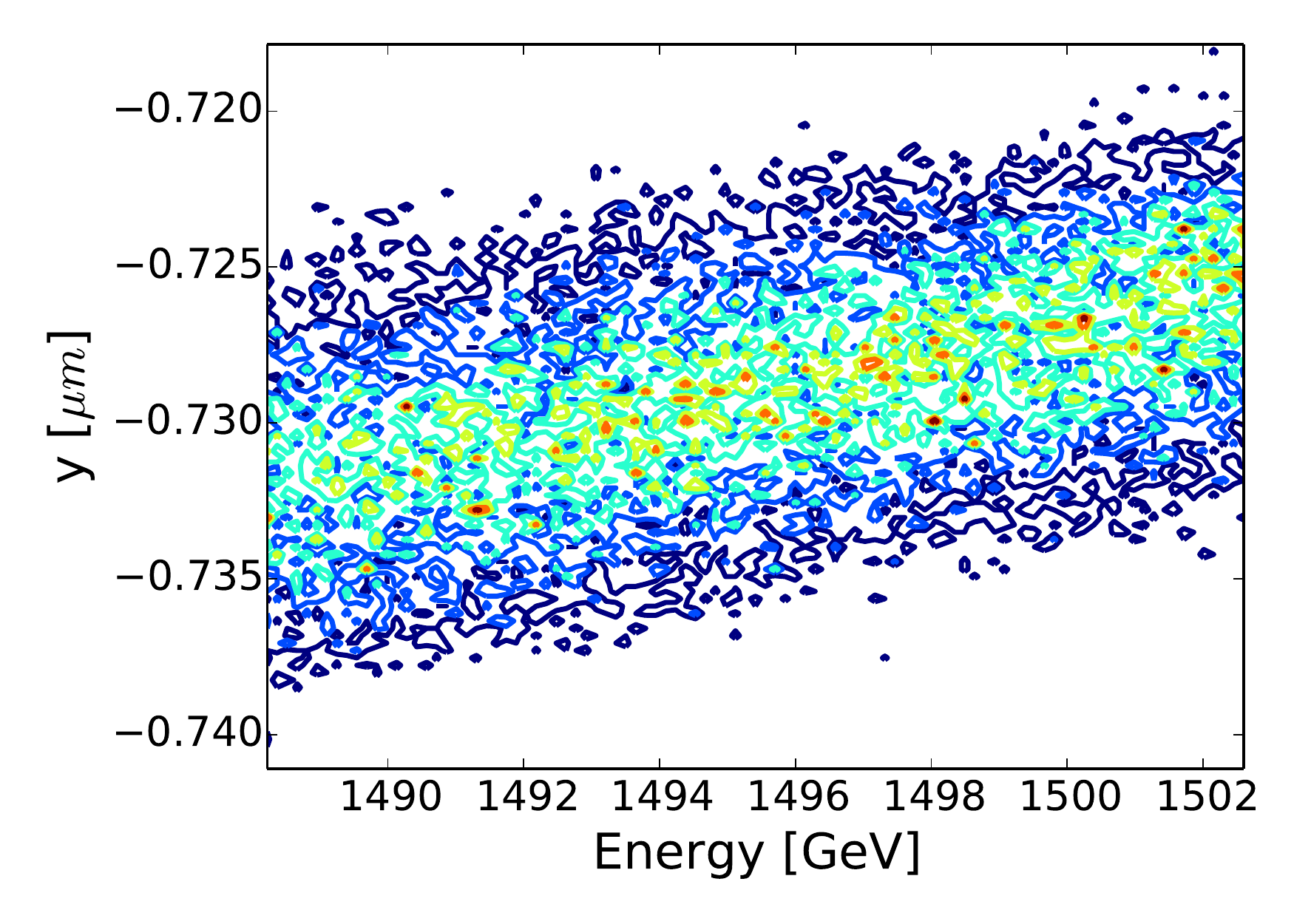}
   % contour_Ey.eps: 0x0 pixel, 300dpi, 0.00x0.00 cm, bb=54 216 558 576
   \label{fig:contour_Ey_wsolas}
 }
\caption{The vertical dispersion with and without solenoid field. No coupling is present in the baseline \subref{fig:contour_Ey_nsol}. The 
solenoid alone introduces a strong coupling \subref{fig:contour_Ey_wsol}, most of which is corrected by the anti-solenoid 
\subref{fig:contour_Ey_wsolas}. Additionally, the solenoid is producing a strong orbit deflection. In \subref{fig:contour_Ey_wsol} the
average vertical position is \unit{6.4}{\micro\meter} off centre.}
\label{fig:contour_Ey}
\end{figure*}

\begin{figure*}
\centering
 \subfigure[~No solenoid]{
   \includegraphics[width=0.45\textwidth]{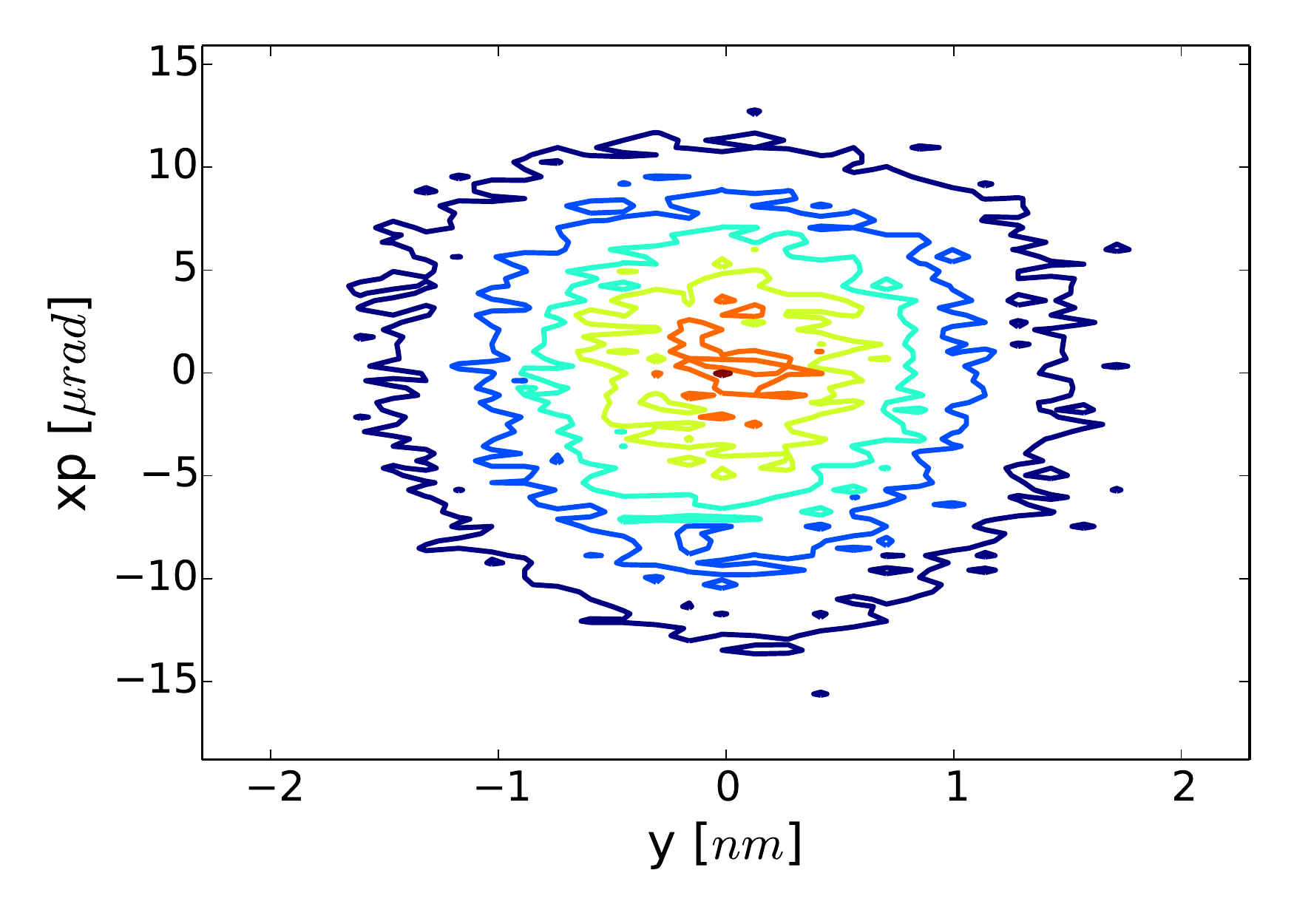}
   \label{fig:contour_yxp_nsol}
 }
 \subfigure[~With solenoid]{
   \includegraphics[width=0.45\textwidth]{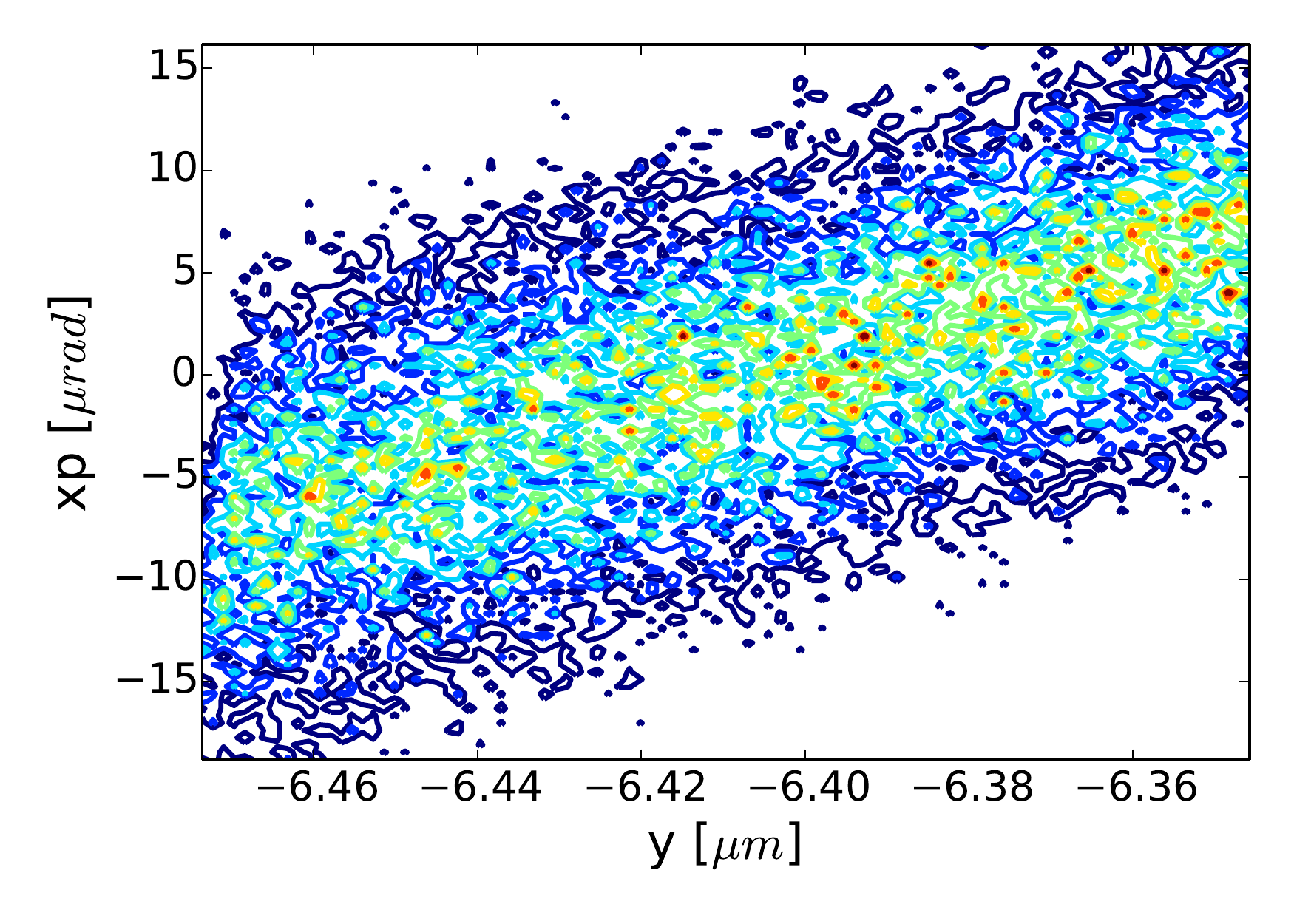}
   \label{fig:contour_yxp_wsol}
 }
 \subfigure[~With solenoid+anti-solenoid]{
   \includegraphics[width=0.45\textwidth]{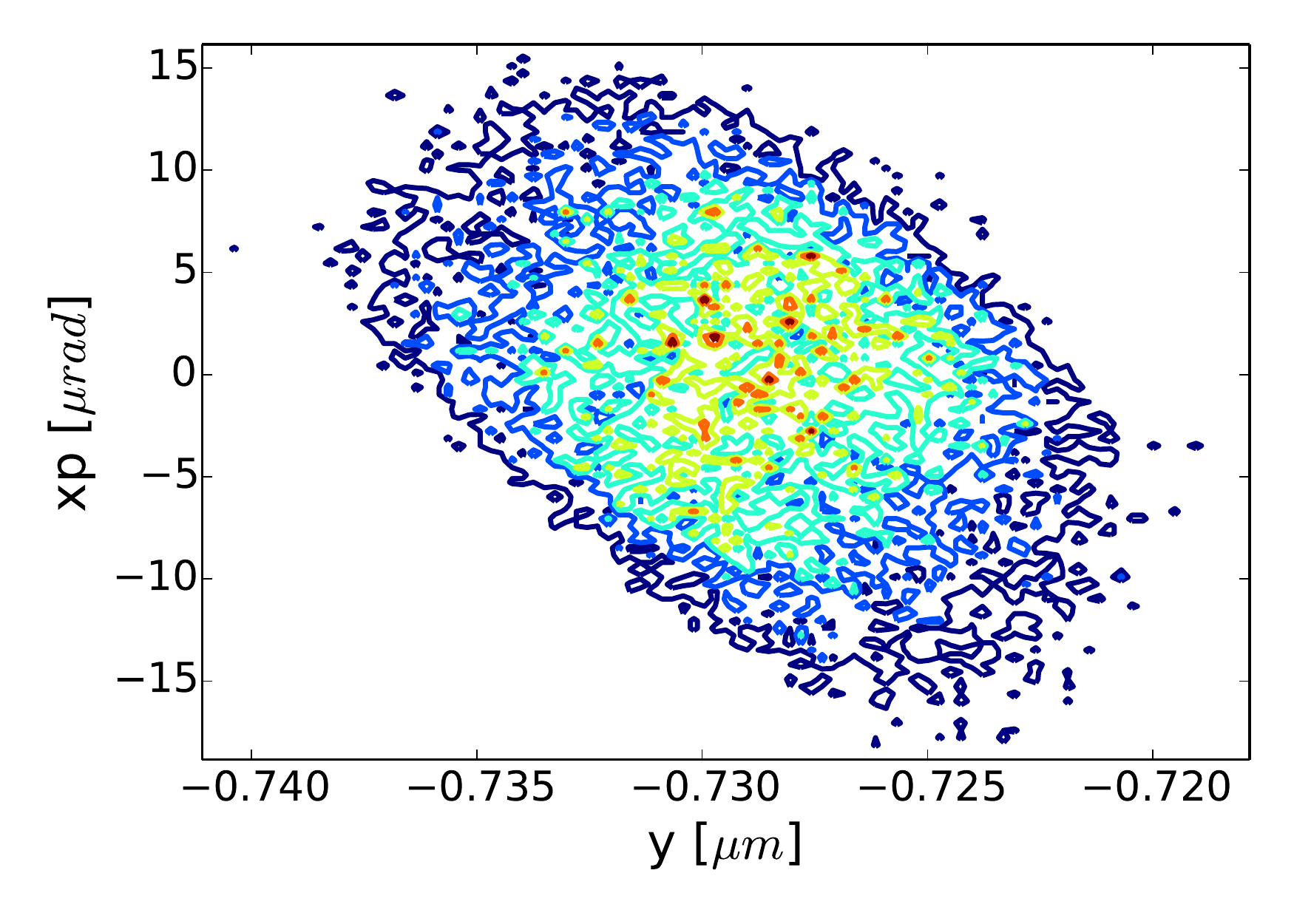}
   \label{fig:contour_yxp_wsolas}
 }
\caption{The y-x$'$ coupling with and without solenoid field. The observations are similar to those observed for the vertical dispersion in 
Fig.~\ref{fig:contour_Ey}.}
\label{fig:contour_yxp}
\end{figure*}

The main distortions responsible for the luminosity loss are vertical dispersion and y-x$'$ coupling, shown in 
Figs.~\ref{fig:contour_Ey}~and~\ref{fig:contour_yxp}.
From these results it is evident that the anti-solenoid alone is not able to
fully compensate the optical distortions caused by the main solenoid field. With the solenoid alone on the order of 1\% of nominal 
luminosity remains, before any compensation of the beam distribution is applied. When we add the anti-solenoid the luminosity increases by 
an order of magnitude, but is still far off acceptable performance.
Other compensation methods are required in addition to the anti-solenoid, in order to fully recover the luminosity.

In order to recover the residual optical distortions induced by the main solenoid field we can use knobs based on transversal sextupole 
displacement in addition to the anti-solenoid. These linear combinations of displacements of the five sextupoles in 
the CLIC final focus system ideally give 5 degrees of freedom for correcting coupling and dispersion terms (vertical displacements),
and 5 for correcting focusing and dispersion terms (horizontal displacements). The main couplings caused by the experimental solenoid are 
vertical dispersion and y-x$'$ coupling, both corrected by vertical knobs. The knobs have been proven successful when applied to 
the tuning against magnet misalignment of the CLIC final focus, as reported in~\cite{Dalena12:BDStuninluminmonitCLIC}.
Additionally, a vertical displacement of QD0 is effective at correcting the vertical offset and dispersion
at the IP. We also add horizontal displacement and roll of the QD0, for more local corrections. With the same three knobs for QF1, we have 
a total of 17 knobs to recover the residual luminosity loss due to the given experimental solenoid and anti-solenoid design.

In our simulation, each knob is evaluated separately. We start with the QD0 knobs, then QF1, then the vertical 
sextupole knobs, and finally the horizontal knobs. For each knob we make a parabolic fit of the luminosity as a 
function of the knob value, and move the magnets accordingly. This is repeated with smaller and smaller steps to make sure we 
are close to the optimum.
We iterate over the entire algorithm a few times to make sure we have at least found a local optimum.
Note that we have not taken into account any magnet imperfections or misalignments in these simulations. ISR is still activated, which 
means that we expect to reach a luminosity of about 96~\% with the anti-solenoid compared to the beamline without solenoid field included, 
based on the result from the deterministic simulation.

\begin{figure}
 \centering
 \includegraphics[width=0.5\textwidth]{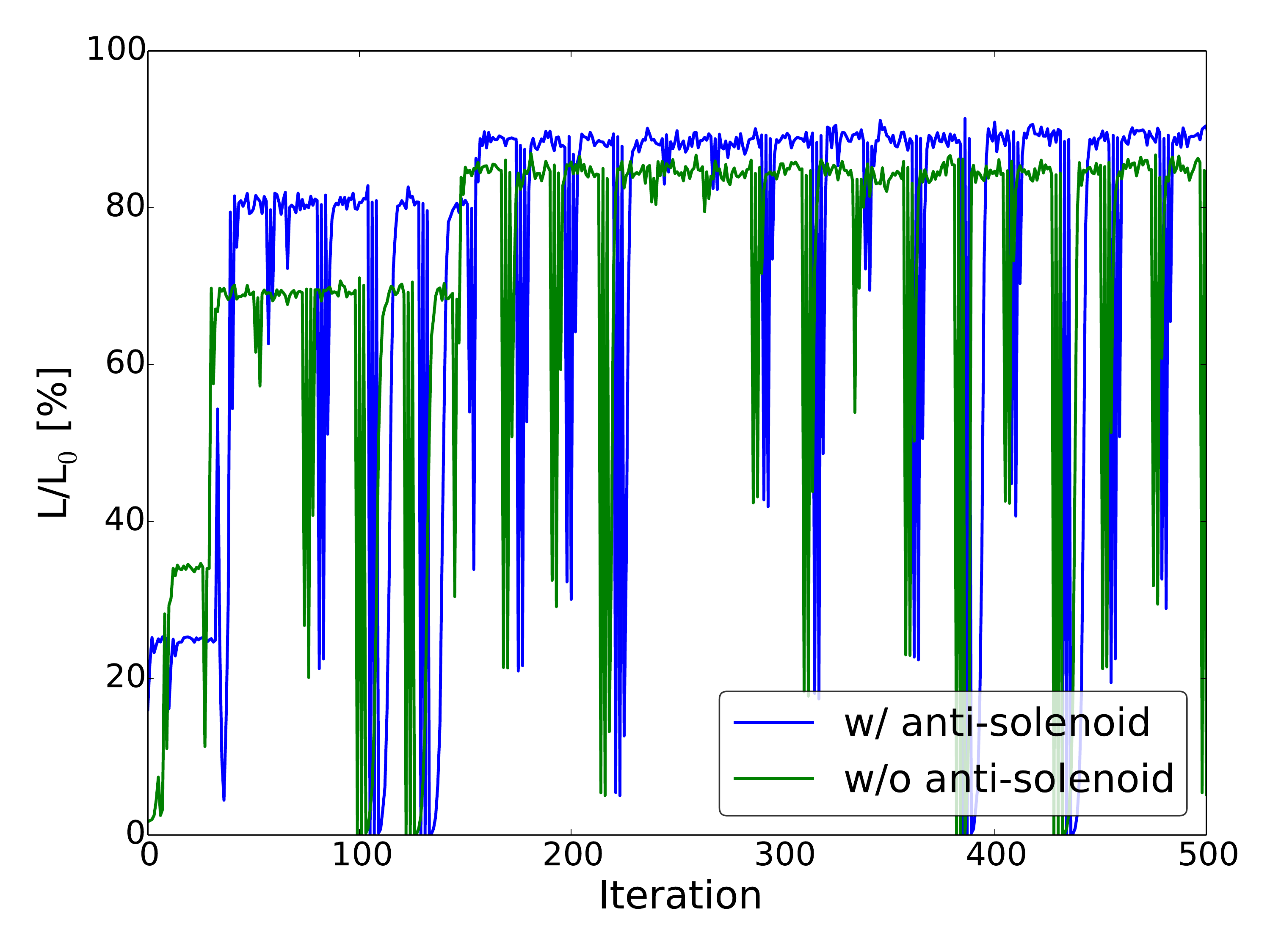}
 % lumi_iterations.pdf: 864x648 pixel, 72dpi, 30.48x22.86 cm, bb=0 0 864 648
 \caption{The luminosity as a function of the number of iterations. The results when including the 
anti-solenoid (blue) is significantly better than the results without (green). 100~\% is defined as the luminosity without solenoid field.}
 \label{fig:lumi_iterations}
\end{figure}

In Fig.~\ref{fig:lumi_iterations} we see the resulting luminosity as a function of iterations. Each dip corresponds to the iteration 
where the algorithm moved to a new knob. We see that most of the aberrations are corrected after the first round of QD0 and vertical 
sextupole knobs. Without the anti-solenoid, it is not possible to obtain the same luminosity level. The number of iterations to reach 
optimal luminosity is about the same.

\begin{figure}
 \centering
  \subfigure[~Vertical dispersion]{
    \includegraphics[width=0.45\textwidth]{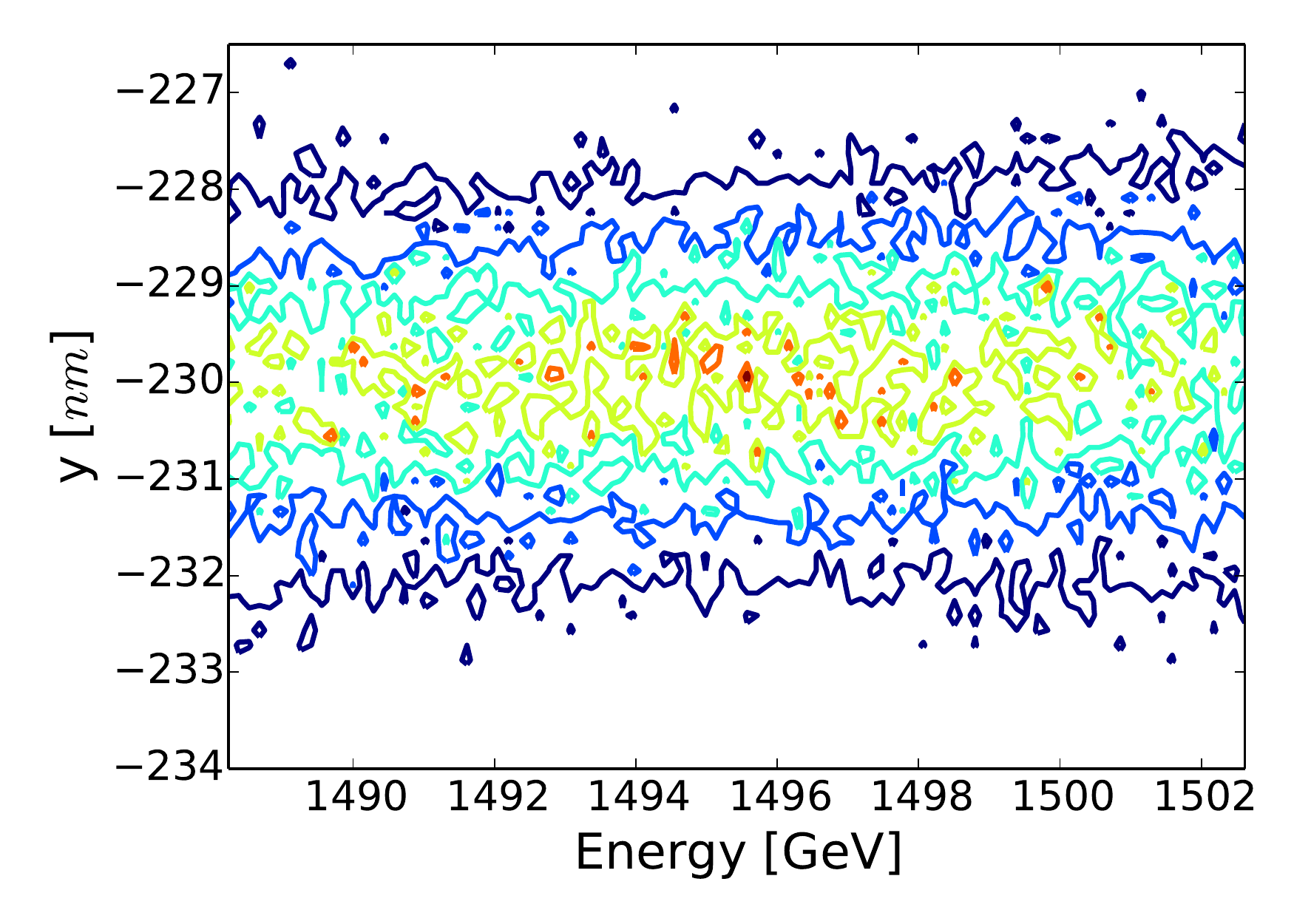}
    % contour_Ey.eps: 0x0 pixel, 300dpi, 0.00x0.00 cm, bb=54 216 558 576
  }
  \subfigure[~y-x$'$ coupling]{
    \includegraphics[width=0.45\textwidth]{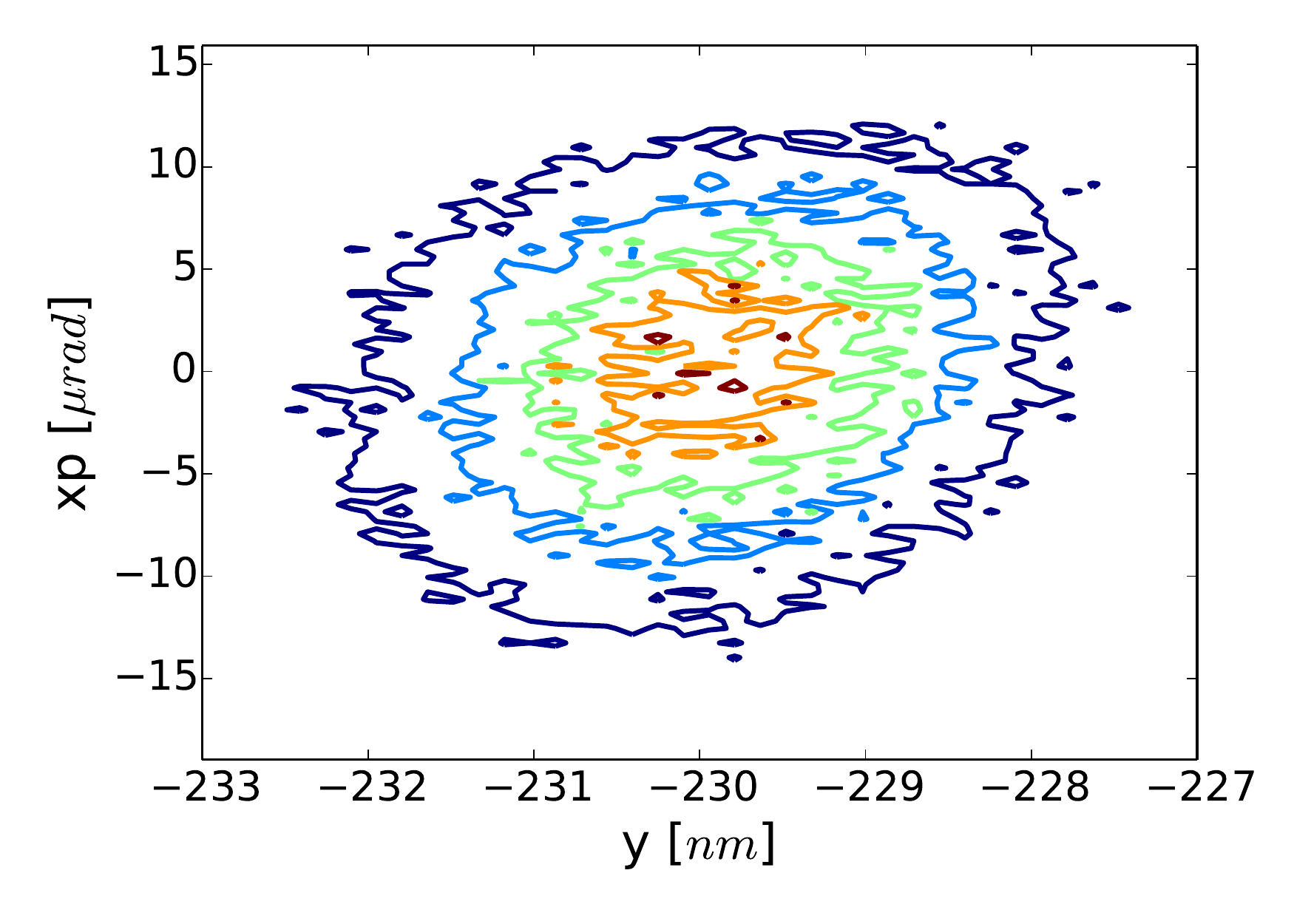}
    % contour_Ey.eps: 0x0 pixel, 300dpi, 0.00x0.00 cm, bb=54 216 558 576
  }
 \caption{The vertical dispersion and y-x$'$ coupling with the optimal knobs. This should be compared to the reference distribution
shown in Figs. \ref{fig:contour_Ey_nsol} and \ref{fig:contour_yxp_nsol}.}
 \label{fig:contour_aftertune}
\end{figure}

We find a luminosity loss of ($8.0\pm 1.6$)~\% with the anti-solenoid in these simulations.
Including the error bars, this fits well with both the semi-analytical estimate of 6~\%, and the deterministic approach which 
estimated 4~\% luminosity 
loss. The results give us confidence that the deterministic approach makes valid assumptions for evaluating the luminosity loss due to ISR 
alone.

In Fig.~\ref{fig:contour_aftertune} the vertical dispersion and y-x$'$ coupling can be seen with the full compensation. Only 
dipolar and quadrupolar terms have been used for this compensation. A check was made keeping ISR off for the entire simulation. The routine 
then completely cancelled out the optical aberrations using only these linear elements. While one can then conclude that the solenoid 
itself introduces mostly linear coupling terms, the solenoid in combination with ISR can lead to non-linear effects that may require 
non-linear correctors. Indeed, the limitation of a tuning-based algorithm is the lack of knowledge about the absolute optimal luminosity.

\section{Conclusions}

A novel simulation approach for estimating the irreversible luminosity loss from incoherent synchrotron radiation produced by the 
experimental solenoid in a high energy lepton collider has been developed. The results are compatible with the slower and more complicated 
simulations to find the full compensation, and consistent with a semi-analytical estimate of the beam size growth. This method obtains
in a deterministic way the optimal luminosity that can be achieved if the correction is perfect.

For the current SiD design for CLIC, we find that we can expect a luminosity loss due to incoherent synchrotron radiation of ($4.1 \pm 
0.2$)~\%. 
This is at the optimistic end of the scale given in the conceptual design report, and is a promising result for the CLIC design 
effort. The anti-solenoid reduces the losses due to ISR by approximately 1~\%, and strongly reduces the optical 
distortions. We have shown through a full simulation that the beam delivery system provides enough flexibility to correct for the 
optical distortions introduced by the solenoid and its overlap with the last focusing magnet.

\section{Acknowledgements}
The authors would like to express their thanks to Michele Modena, Hubert Gerwig, Antonio Bartalesi, and Alexander Aloev for all their help 
with the field map simulations and fruitful discussions. We would also like to thank Andrea Latina and Jochem Snuverink for valuable help 
implementing the code into PLACET. Finally, we would like to thank Paul Anton Letnes for carefully reading through and correcting the 
English grammar.

\section{APPENDIX: Tracking Routine}

We have implemented a 4th order symplectic integrator in the particle tracking code PLACET 
\cite{LatiAdliBurkReniRumoSchuToma08:RecenImproTrackCodePLACE, Latina12:PLACETimprovements}, with a user defined step length.
The sum of magnetic fields from beamline elements and solenoid field map is used to calculate the Lorentz force at each location of a kick. 
With the appropriate choices for drifts and kicks, it can be shown numerically that this integrator is in fact of 4th order 
\cite{Forest90:Fourtsymplinteg}. This integrator allows us to track the beam through a combination of beamline elements and added field map 
(solenoid field), something that was not possible previously.

The new integrator has been compared to the other independent tracking routines already available in 
PLACET (excluding the solenoid field map), and found to be in good agreement. The 4th order integrator has also been independently compared 
to a Lie tracking routine~\cite{Dalena14:fringeFieldsModeling}.

\begin{figure}
 \centering
 \includegraphics[width=0.5\textwidth]{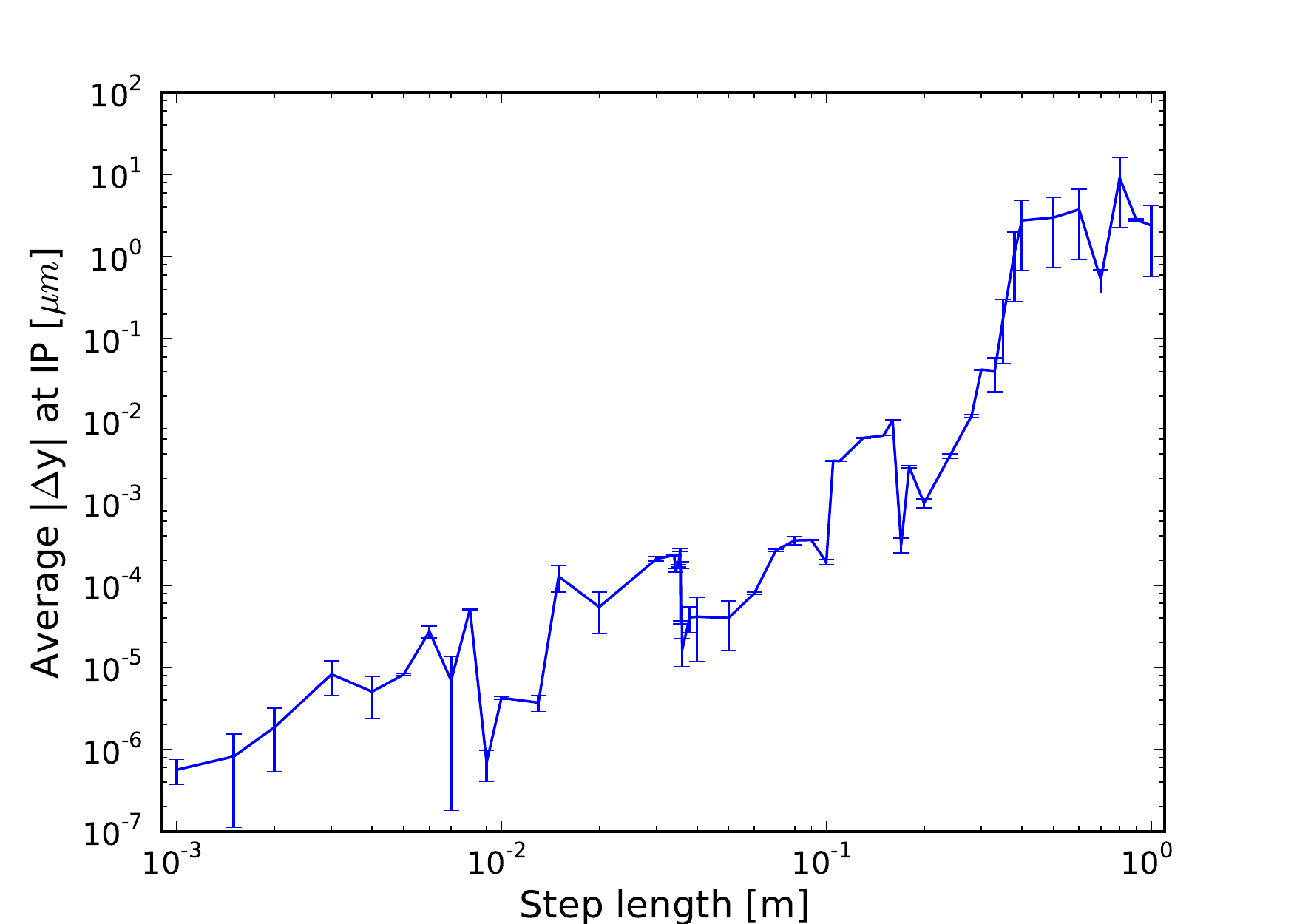}
 % step_length_error.pdf: 504x360 pixel, 72dpi, 17.78x12.70 cm, bb=0 0 504 360
 \caption{The difference in vertical position as a function of step length is evaluated for 4000 initial positions, using the 
formulae defined in Eq.~\eqref{eq:avgdeltay}. The trajectory with shortest step length (\unit{0.2}{\milli\meter}) is used as 
reference.}
 \label{fig:step_length_error}
\end{figure}

In Fig.~\ref{fig:step_length_error} the error in the vertical position at the interaction point is shown as a function of the step 
length
used. Each set of initial coordinates is tracked with multiple step lengths. The error is estimated as the final position at the
interaction point with the
given step length compared to using a much shorter step length. For each step length we then get an average error for $N$ initial
coordinates as 
\begin{equation}
\text{avg}(|\Delta y|)= \frac{\sum_{i=1}^N{|\Delta y_i|}}{N}.
\label{eq:avgdeltay}
\end{equation}

The vertical beam size at the interaction point is approximately \unit{1}{\nano\meter}, so the error should be well below this value.
Hence, step lengths lower than \unit{1}{\centi\meter} are acceptable. The results presented in this paper are obtained using a step
length of \unit{1}{\milli\meter}.

\bibliography{references}
\end{document}